\documentclass{PoS}
\usepackage{amsmath}

\title{
\begin{picture}(0,0)(0,0)%
   \put(350,60){\makebox(0,0)[l]{\textnormal{\normalsize 
   J-PARC-TH-0211 }}}%
   \end{picture}%
Classifying topological sector via machine learning
}

\ShortTitle{Classifying topological sector via machine learning}

\author{\speaker{Masakiyo Kitazawa}\\
  Department of Physics, Osaka University, Toyonaka, Osaka 560-0043, Japan \\
  J-PARC Branch, KEK Theory Center, Institute of Particle and Nuclear Studies, KEK, 203-1, Shirakata, Tokai, Ibaraki, 319-1106, Japan\\
  E-mail: \email{kitazawa@phys.sci.osaka-u.c.jp}
}

\author{Takuya Matsumoto\\
  Department of Physics, Osaka University, Toyonaka, Osaka 560-0043, Japan
}

\author{Yasuhiro Kohno\\
  Research Center for Nuclear Physics, Osaka University, Ibaraki, Osaka 567-0047, Japan
}

\abstract{
  We employ a machine learning technique for an estimate of the
  topological charge $Q$ of gauge configurations in SU(3) Yang-Mills
  theory in vacuum~\cite{Matsumoto:2019jia}.
  As a first trial, we feed the four-dimensional topological charge
  density with and without smoothing into the convolutional neural
  network and train it to estimate the value of $Q$.
  We find that the trained neural network can estimate the 
  value of $Q$ from the topological charge density at small flow time
  with high accuracy.
  Next, we perform the dimensional reduction of the input data
  as a preprocessing and analyze lower dimensional data by the neural network.
  We find that the accuracy of the neural network
  does not have statistically-significant dependence on the dimension of
  the input data.
  From this result we argue that the neural network does not find
  characteristic features responsible for the determination of $Q$
  in the higher dimensional space.
}

\FullConference{37th International Symposium on Lattice Field Theory - Lattice2019\\
		16-22 June 2019\\
		Wuhan, China}

\begin{document}

\section{Introduction}

The existence of nontrivial topology is one of the most important
non-perturbative aspects of Quantum chromodynamics (QCD) and other
Yang-Mills (YM) gauge theories in four spacetime dimensions.
These theories can have topologically nontrivial
gauge configurations classified by the topological charge $Q$.
The topology in QCD is responsible for various non-perturbative
properties of this theory, such as the U(1) problem~\cite{Weinberg:1996kr}.

The topological property of YM theories has been
studied by numerical simulations of lattice gauge theory.
Although gauge configurations on the lattice are strictly speaking
topologically trivial, it is known that well separated topological
sectors emerge as the continuum limit is approached~\cite{Luscher:1981zq}.
It is known that the values of $Q$ of lattice gauge configurations
measured by various methods show an approximate
agreement~\cite{Alexandrou:2017hqw}, which is consistent with 
the existence of the well separated topological sectors
in lattice gauge theory.

In the present study, we apply the machine learning (ML) 
for the analysis of $Q$ of gauge configurations on the
lattice~\cite{Matsumoto:2019jia}.
We generate data by the numerical simulation of SU(3) YM theory 
in four spacetime dimensions, and feed them into the neural networks (NN).
The main motivation of this study is the search for 
characteristic local structures in the four-dimensional space
related to $Q$ by the ML.
It is known that YM theories have classical gauge configurations
called instantons, which carry a nonzero topological charge and 
have a localized structure~\cite{Weinberg:1996kr}.
If the topological charge of the quantum gauge configurations
is also carried by instanton-like local objects,
the NN would recognize and make use of them for the prediction of $Q$.
This study will also contribute to a reduction of the numerical costs 
for the analysis of $Q$.

\section{Topological charge and gradient flow}
\label{sec:Q}

In this study, we consider 
SU(3) YM theory in the four-dimensional Euclidean space
with the periodic boundary conditions for all directions.
The Wilson gauge action is used for generating gauge configurations.
The numerical analyses are performed 
at two inverse couplings $\beta=6/g^2=6.2$ and $6.5$ with 
the lattice volumes $16^4$ and $24^4$, respectively, and
$20,000$ gauge configurations have been generated for each analysis.
These two lattices have almost the same physical volume.

\begin{figure}[t]
  \centering
  \begin{minipage}{0.49\textwidth}
  \includegraphics[width=0.99\textwidth,clip]{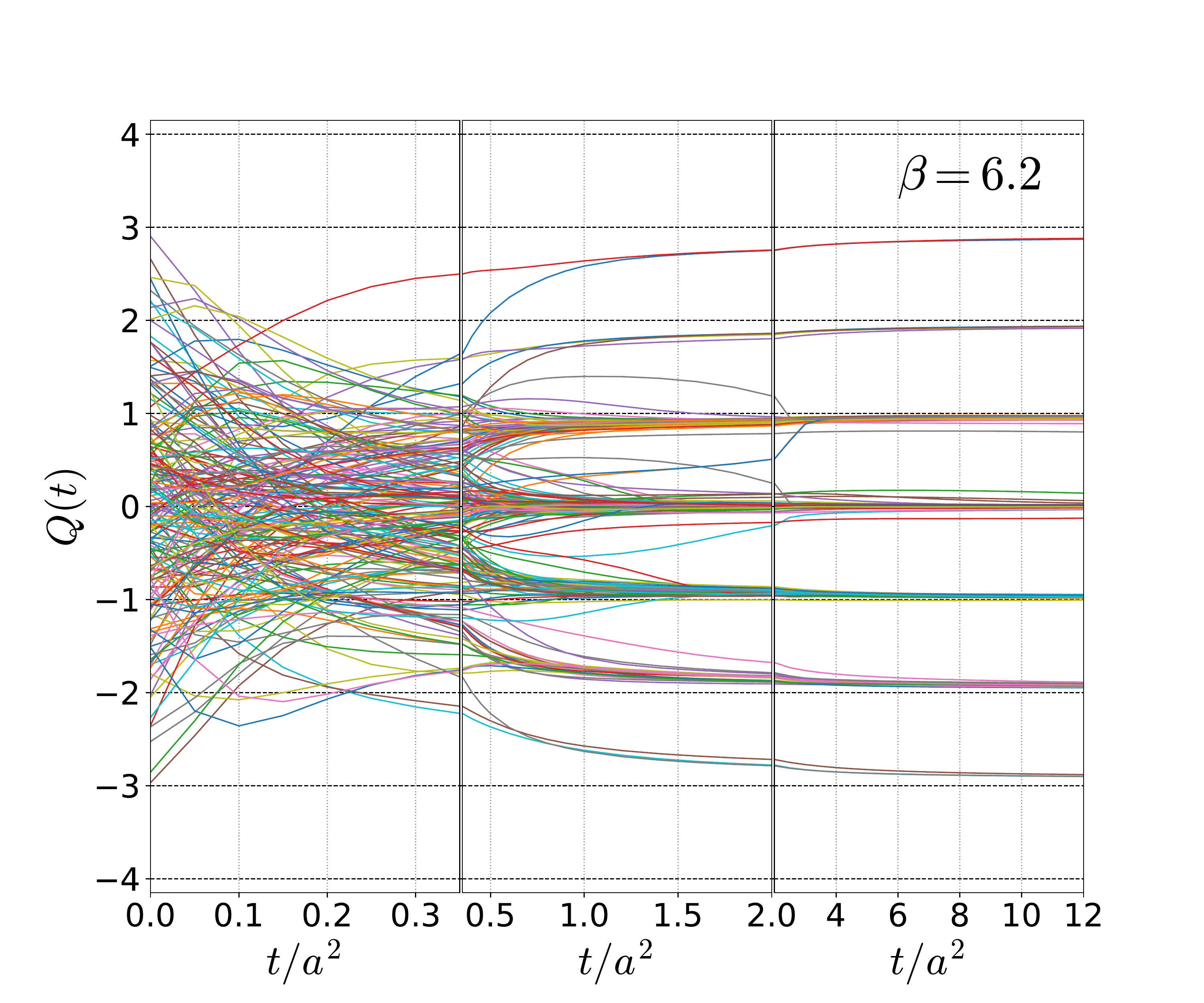}
  \includegraphics[width=0.99\textwidth,clip]{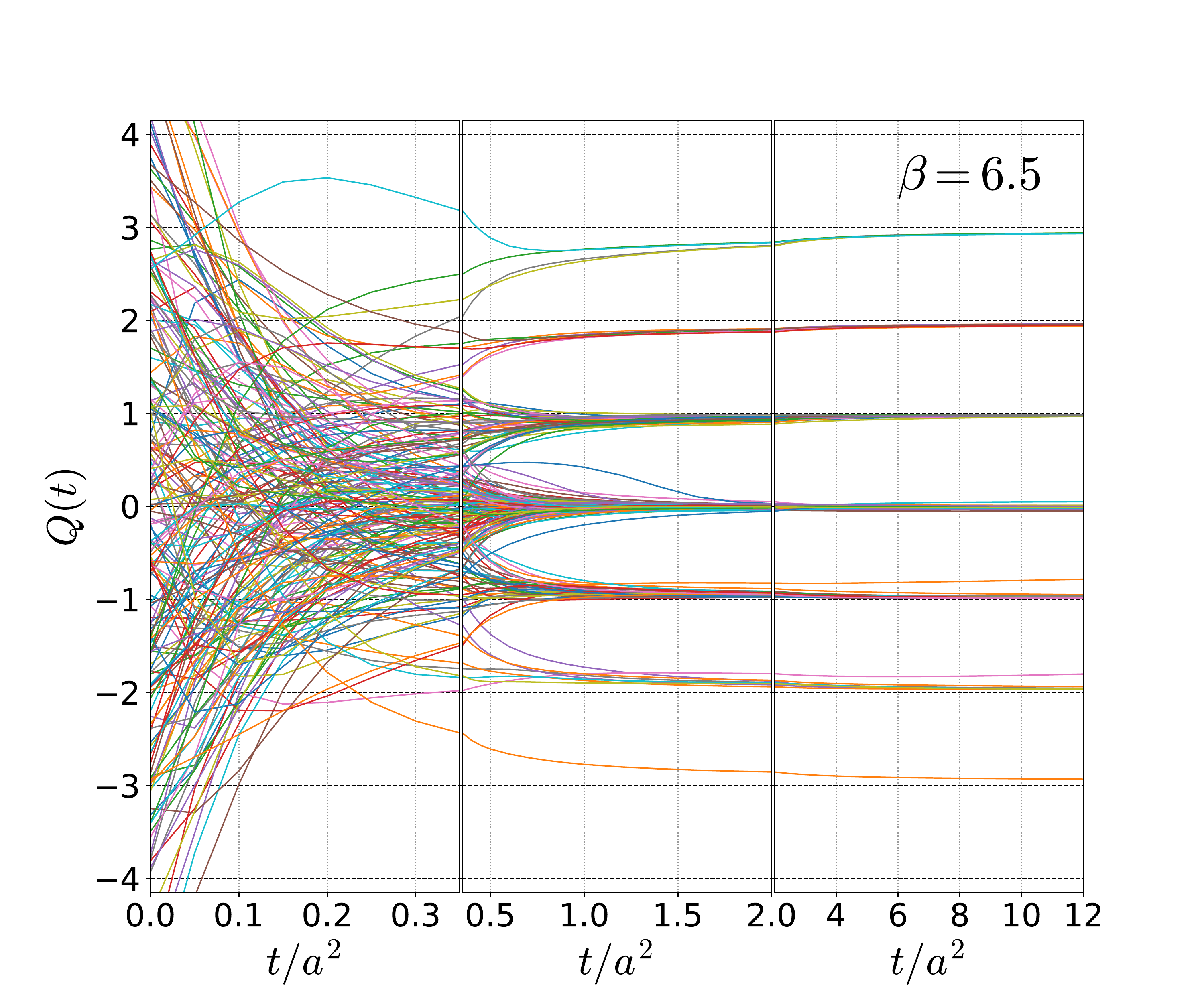}
  \end{minipage}
  \begin{minipage}{0.49\textwidth}
    \vspace{4mm}
  \includegraphics[width=0.9\textwidth,clip]{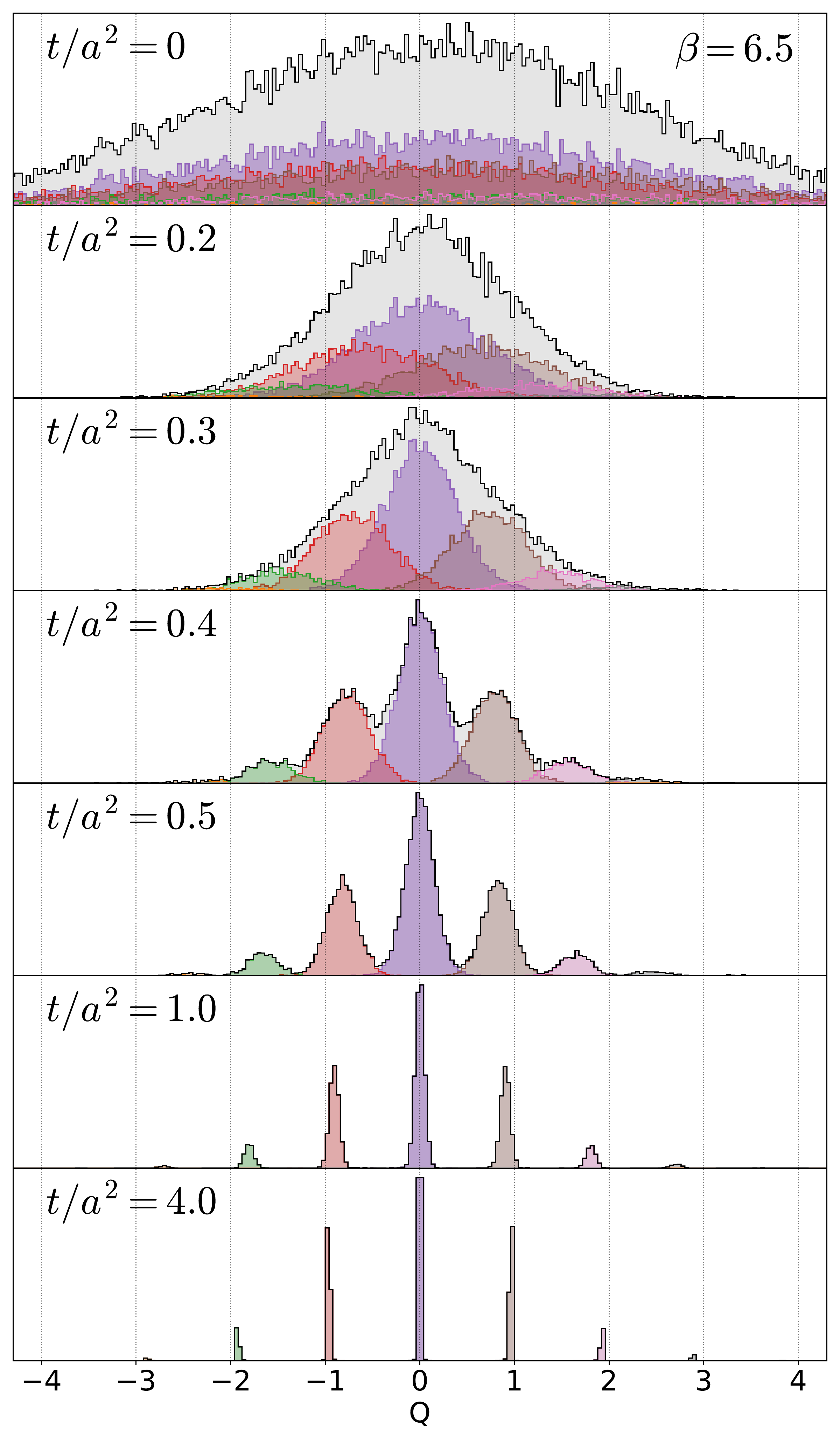}
  \end{minipage}
  \caption{
    Left:
    Flow time $t$ dependence of $Q(t)$ on $200$ gauge configurations
    at $\beta=6.2$ (upper) and $6.5$ (lower).
    Right:
    Distribution of $Q(t)$ at several values of $t/a^2$.
    The colored histograms are the distributions in individual
    topological sectors.
  }
  \label{fig:Q(t)}
\end{figure}

In the continuous YM theory
the topological charge is defined by
\begin{align}
  Q = \int_V d^4x\,q(x),
  \label{eq:Q}
  \qquad
  q(x) = -\frac{1}{32\pi^2}\epsilon_{\mu\nu\rho\sigma}
  {\rm tr} \left[F_{\mu\nu}(x)F_{\rho\sigma}(x)\right],
\end{align}
with
the field strength
$F_{\mu\nu}(x)=\partial_\mu A_\nu(x)-\partial_\nu A_\mu(x)+[A_\mu(x),A_\nu(x)]$
and $q(x)$ the topological-charge density.
In lattice gauge theory, Eq.~(\ref{eq:Q}) calculated on a gauge
configuration is not given by an integer, but distributes continuously.
To obtain discretized values, one may apply a smoothing
of the gauge field before the measurement of $q(x)$.
In the present study, we use the gradient
flow~\cite{Luscher:2011bx} for the smoothing.
The gradient flow is a transformation of the gauge field
described by a continuous parameter $t$ called the flow time.
The gauge field at a flow time $t$ is a smoothed
field with the mean-square smoothing radius
$\sqrt{8t}$~\cite{Luscher:2011bx}.
We denote the topological charge density at $t$ as 
$q_t(x)$, and its four-dimensional integral as
\begin{align}
  Q(t) = \int_V d^4 x  \, q_t(x). 
\end{align}

In the left panels of Fig.~\ref{fig:Q(t)}, we show
$200$ examples of $Q(t)$ for $\beta=6.2$ and $6.5$.
The horizontal axis shows $t/a^2$ with the lattice spacing $a$.
The panels show that $Q(t)$ approaches discrete integer values
as $t$ becomes larger.
In the right panel of Fig.~\ref{fig:Q(t)}, we show the distribution of
$Q(t)$ at $\beta=6.5$
for several values of $t/a^2$ by the histogram.
Although the values of $Q(t)$ are distributed continuously around the origin
at $t=0$, the distribution converges on discretized integer values
as $t$ becomes larger.
At $t/a^2\gtrsim1.0$, the gauge configurations are well separated
into different topological sectors.
We define the topological charge from $Q(t)$
at $t/a^2=4.0$ as 
\begin{align}
  Q={\rm round}[Q(t)]_{t/a^2=4.0},
  \label{eq:Qdef}
\end{align}
where ${\rm round}(x)$ means rounding off to the nearest integer.
As indicated from Fig.~\ref{fig:Q(t)}, the value of $Q$ in Eq.~(\ref{eq:Qdef})
hardly changes with the variation of $t/a^2$ in the range $t/a^2\gtrsim4$.
It is known that the topological charge $Q$ defined in this way
approximately agrees with those obtained through
other definitions~\cite{Alexandrou:2017hqw}.
In the right panel of Fig.~\ref{fig:Q(t)}, the distributions of $Q(t)$ in
individual topological sectors are shown
by the colored histograms.
From the panel one sees that the
distribution of $Q(t)$ deviates from integer values toward the origin.
This deviation becomes smaller as $t$ becomes larger.

\section{Benchmark}
\label{sec:bench}

In this study, we evaluate the performance of a 
model for an estimate of $Q$ by the accuracy $P$ and recalls $R_Q$
defined by 
\begin{align}
  P = \frac{N_{\rm correct}}{N_{\rm total}},
  \qquad
  R_Q = \frac{N_Q^{\rm correct}}{N_Q} ,
  \label{eq:R}
\end{align}
where $N_{\rm total}$, $N_{\rm correct}$, $N_Q$, and
$N_Q^{\rm correct}$ mean 
the numbers of total data, total correct answers,
data in the topological sector $Q$, and
correct answers among them, respectively.
The recalls are suitable to see the bias of the answers.

\begin{table}
   \centering
   \scalebox{0.8}{
   \begin{tabular}{c|cc}
     \hline
     \hline
     $t/a^2$ & $\beta=6.2$ & $\beta=6.5$   \\
     \hline
     0   & 0.273(3) & 0.162(3) \\
     0.1 & 0.383(4) & 0.274(3) \\
     0.2 & 0.546(4) & 0.474(4) \\
     0.3 & 0.773(3) & 0.713(4) \\
     0.4 & 0.925(2) & 0.916(2) \\
     \hline
     \hline
   \end{tabular}
   \hspace{1cm}
   \begin{tabular}{c|cc}
     \hline
     \hline
     $t/a^2$ & $\beta=6.2$ & $\beta=6.5$   \\
     \hline
     0.5 & 0.960(1) & 0.989(1) \\
     1.0 & 0.982(1) & 0.999(0) \\
     2.0 & 0.992(1) & 0.999(0) \\
     4.0 & 1.000(0) & 1.000(0) \\
     10.0& 0.993(1) & 0.999(0) \\
     \hline
     \hline
   \end{tabular}
   }   
 \caption{
   Accuracy $P_{\rm imp}$ obtained by the model Eq.~(\ref{eq:Qimp})
   with the optimized parameter $c$.
 }
 \label{table:bench}
\end{table}

In this study, we analyze $q_t(x)$ or $Q(t)$ at small $t$
by the NN.
Here, $t$ used for the input has to be chosen small enough
so that a simple estimate of $Q$ like Eq.~(\ref{eq:Qdef})
\begin{align}
  Q_{\rm naive}={\rm round} [Q(t)],
  \label{eq:Qnaive}
\end{align}
does not give a high accuracy.
We also consider an improved estimator of $Q$ defined by
\begin{align}
  Q_{\rm imp}={\rm round}[ cQ(t)] , 
  \label{eq:Qimp}
\end{align}
where $c$ is a parameter determined so as to maximize the accuracy
in the range $c>1$ for each $t$.
This parameter is introduced to compensate the deviation of the
distribution of $Q(t)$ in each topological sector from integers
found in Fig.~\ref{fig:Q(t)}.
We found that Eq.~(\ref{eq:Qimp}) with the optimized $c$ has a better
accuracy compared with the naive model Eq.~(\ref{eq:Qnaive}) at 
some range of $t/a^2$.
In Table~\ref{table:bench}, we show the accuracy $P_{\rm imp}$
of the model Eq.~(\ref{eq:Qimp}) for several values of $t/a^2$.
We note that $P_{\rm naive}=P_{\rm imp}=1$ at $t/a^2=4.0$ by definition.
As the accuracy is almost unity at $t/a^2=2.0$ and $10.0$,
the value of $Q$ defined by Eq.~(\ref{eq:Qdef}) hardly changes with
the variation of $t/a^2$ in the range $t/a^2\gtrsim2.0$.

\section{Analysis of four-dimensional field}
\label{sec:4}

\begin{table}
   \centering
   \scalebox{0.8}{
   \begin{tabular}{c|cc}
     \hline
     \hline
     layer & filter size & output size \\
     \hline
     input       & -     & $8^d\times N_{\rm ch}$ \\
     convolution & $3^d$ & $8^d\times 5$ \\
     convolution & $3^d$ & $8^d\times 5$ \\
     convolution & $3^d$ & $8^d\times 5$ \\
     global average pooling & $8^d$ & $1\times5$ \\
     full connect & - & 5 \\
     full connect & - & 1 \\
     \hline
     \hline
   \end{tabular}
   \hspace{1cm}
   \begin{tabular}{c|c}
     \hline
     \hline
     layer & output size \\
     \hline
     input        & 3 \\
     full connect & 5 \\
     full connect & 1 \\
     \hline
     \hline
   \end{tabular}
   }
   \caption{
     Designs of the NN.
     In Secs.~\ref{sec:4} and \ref{sec:dim_red} we use the convolutional
     NN in the left panel with 
     the dimension $d$ of the input data, while in Sec.~\ref{sec:0}
     a simple fully-connected NN in the right panel is employed.
   }
   \label{table:network}
\end{table}

In this section, we analyze the four-dimensional data of $q_t(x)$
by the ML.
We feed the topological charge density $q_t(x)$ 
into the NN and train it by the supervised learning
so that it answers the value of $Q$ defined by Eq.~(\ref{eq:Qdef}).
We employ the convolutional NN (CNN) for four-dimensional space.
As the CNN is a class of NNs which had been developed for the image
recognition~\cite{726791,Krizhevsky:2012},
this framework would be suitable for searching for 
local features in the four-dimensional data.

The structure of the CNN is shown in the left panel of Table~\ref{table:network}.
The lattice volume is reduced to $8^4$ from $16^4$ and $24^4$
by the average pooling by a preprocessing.
The CNN has three convolutional layers with the filter size $3^4$ and
five output channels.
The global average pooling (GAP) layer, which takes the average
with respect to the spatial coordinates, is inserted after the
convolutional layers to respects the translational symmetry of the input data.
The output of the GAP layer is then processed by two fully-connected
layers.
For more details on the NN and the procedure of the supervised learning,
see Ref.~\cite{Matsumoto:2019jia}.

In the upper four rows of Table~\ref{table:4single}, we show
the resulting accuracy $P$ and recall of each topological sector $R_Q$
with the input data at several flow times $t$ for $\beta=6.5$.
At $t=0$, the accuracy takes a nonzero value $P\simeq0.39$.
However, from the values of $R_Q$
one finds that the NN answers $Q=0$ for almost all gauge configurations.
This result shows that the CNN does not find any useful
features in the input data.

\begin{table}
  \centering
  \scalebox{0.8}{
  \begin{tabular}{c|c|c|ccccccccc}
     \hline
     \hline
     & & & \multicolumn{9}{c}{$R_Q$}
     \\
     \cline{4-12}
     $N_{\rm ch}$ & input $t/a^2$ & $P$
     & -4 & -3 & -2 & -1 & 0 & 1 & 2 & 3 & 4 
     \\
     \hline
     1 & 0 & 0.388    & 0 & 0 & 0 & 0 & 1.000 & 0 & 0 & 0 & 0
     \\
     1 & 0.1 & 0.396  & 0 & 0 & 0 & 0.086 & 0.889 & 0.129 & 0 & 0 & 0
     \\
     1 & 0.2 & 0.479  & 0 & 0 & 0.108 & 0.445 & 0.641 & 0.459 & 0.150 & 0 & 0
     \\
     1 & 0.3 & 0.698  & 0 & 0.170 & 0.585 & 0.730 & 0.727 & 0.701 & 0.624 & 0.395 & 0.071  
     \\
     3 & 0.3,0.2,0.1 & 0.953
     & 0 & 0.830 & 0.951 & 0.956 & 0.952 & 0.962 & 0.968 & 0.953 & 0.286
     \\
     \hline
     \hline
   \end{tabular}
   }
 \caption{
   Accuracy $P$ and the recalls of individual topological
   sectors $R_Q$ obtained by the analysis of $q_t(x)$
   in the four-dimensional space by the CNN for $\beta=6.5$.
   The input data has $N_{\rm ch}$ channels.
 }
 \label{table:4single}
\end{table}

Because the analysis of the data at $t=0$ does not provide any
useful results, as a next trial we perform the analysis of $q_t(x)$
at nonzero $t$.
The resulting accuracy and recalls at $t/a^2=0.1$, $0.2$, and $0.3$
depicted in Table~\ref{table:4single} shows
that the accuracy becomes better as $t/a^2$ becomes
larger.
From the behavior of $R_Q$ one also finds that the answers of
the CNN are distributed for all topological sectors at large $t/a^2$.
From this result, it is na\"ively expected that the CNN recognizes
features in the four-dimensional space.
However, by comparing these accuracies with Table~\ref{table:bench},
one finds that the accuracies are almost consistent
with the benchmark model Eq.~(\ref{eq:Qimp}) at the same $t/a^2$.
A natural interpretation of this result is that
the answers of the CNN are obtained by Eq.~(\ref{eq:Qimp}).
Because the four-dimensional structure is completely integrated out
in Eq.~(\ref{eq:Qimp}), this result shows that 
characteristic features responsible for the determination
of $Q$ in the four-dimensional space were not found by the CNN.

In order to realize the recognition of the four-dimensional data 
by the CNN, we next feed $q_t(x)$ obtained at three different
flow times simultaneously as a single multi-channel data.
We use $q_t(x)$ at $t/a^2=0.3$, $0.2$, and $0.1$ for the input.
The resulting accuracy and recalls 
are shown at the bottom row of Table~\ref{table:4single}.
One finds a remarkable improvement of the accuracy
compared with the analysis of the single flow time,
which indicates a nontrivial recognition of the four-dimensional
data by the CNN.

\section{Analysis of $Q(t)$}
\label{sec:0}

Let us inspect whether the high accuracy of the CNN 
in the previous section is established by
the recognition of the four-dimensional space, or not.
For this purpose, in this section we consider a simple fully connected
NN (FNN) which accepts only three values of $Q(t)$ at different $t/a^2$.
The structure of the FNN is shown in the right panel of
Table~\ref{table:network}.
The FNN has only one hidden layer that is fully connected
with the input and output layers.
\begin{table}
  \centering
  \scalebox{0.8}{
   \begin{tabular}{l|cc}
     \hline
     \hline
     input $t/a^2$ & $\beta=6.2$ & $\beta=6.5$   \\
     \hline
     0.35, 0.3, 0.25  & 0.967(2) & 0.996(1) \\
     {\bf 0.3, 0.25, 0.2}   & {\bf 0.959(2)} & {\bf 0.990(2)} \\
     0.25, 0.2, 0.15  & 0.939(3) & 0.951(2) \\
     0.3, 0.2, 0.1    & 0.941(2) & 0.957(2) \\
     \hline
     \hline
   \end{tabular}
  }
   \caption{
     Accuracy of the trained FNN in Table~\ref{table:network}
     with various sets of the input data.
     Left column shows the values of $t/a^2$
     that evaluate $Q(t)$ for the input.
     Errors are estimated from the variance among 10 different trainings.
   }
   \label{table:result0}
\end{table}

In Table~\ref{table:result0}, we show the accuracies obtained by 
the trained FNN for various combinations of the flow times for the input
$Q(t)$ at $\beta=6.2$ and $6.5$.
The table shows that the accuracy with the combination
$t/a^2=(0.3,0.2,0.1)$ is comparable with that obtained in
the previous section, although 
the four-dimensional structure of the input data is fully integrated out
in this analysis.
From this result it is concluded that the CNN in Sec.~\ref{sec:4}
does not find features in the four-dimensional space,
but performs almost the same analysis as the FNN introduced in this section.

Table~\ref{table:result0} also suggests that
the accuracy obtained by the FNN is significantly
higher than $P_{\rm imp}$ with the same largest $t/a^2$.
In particular, the accuracy with 
$t/a^2=(0.3,0.25,0.2)$ shown by the bold letters is 
as high as $99\%$ for $\beta=6.5$, while the benchmark model
Eq.~(\ref{eq:Qimp}) gives $P_{\rm imp}\simeq0.71$ at $t/a^2=0.3$.
This result shows that the trained NN can estimate $Q$ 
quite successfully only with the data at $t/a^2\lesssim0.3$,
and this model can be used for an effective analysis of $Q$.
The robustness of this result against the variation of the lattice
spacing and the effect of the reduction of the training data
are discussed in Ref.~\cite{Matsumoto:2019jia}.

\section{Dimensional reduction}
\label{sec:dim_red}

\begin{figure}[t]
  \centering
  \vspace{5mm}
  \includegraphics[width=0.46\textwidth,clip]{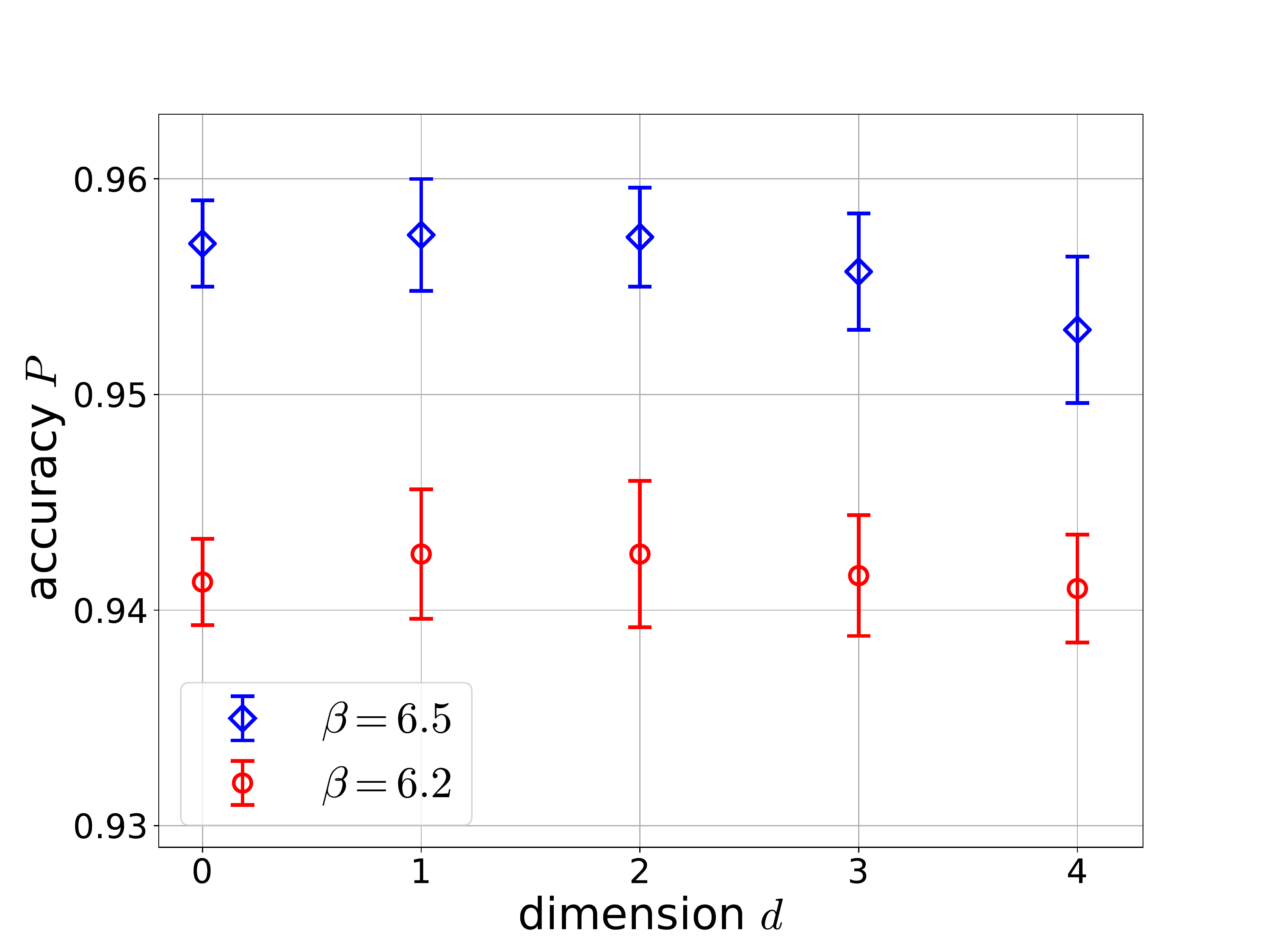}
  \caption{
    Dependence of the accuracy $P$ on the dimension $d$
    of the input data.
  }
  \label{fig:dim_red}
\end{figure}

In Secs.~\ref{sec:4} and \ref{sec:0}, we analyzed the data
with the space dimension $d=4$ and $0$, respectively.
We next consider the analysis of the data with dimensions 
$d=1$--$3$ by the NN.
We reduce the dimension of the input data by the dimensional reduction,
i.e. by integrating out some coordinates, 
and explore a possible optimal dimension of the input data of the CNN
between $d=0$ and $4$.

In Fig.~\ref{fig:dim_red}, we show the dependence of the resulting
accuracy of the trained CNN on $d$ with the flow times of the input data
$t/a^2=(0.3,0.2,0.1)$.
The figure shows that
the accuracy does not have a statistically-significant $d$ dependence,
although the results at $d=1$ and $2$ would be slightly better.
This result supports our observation that the CNN does not recognize
features in the multi dimensional space.

\section{Summary}

In this study, we applied a machine learning technique
for the classification of the topological sector
of gauge configurations in SU(3) YM theory.
We found that the value of $Q$ defined 
at a large flow time can be predicted with high accuracy
only with $Q(t)$ at $t/a^2 \le 0.3$ with the aid of the NN.
This procedure would be used for reducing the numerical cost
for the analysis of $Q$.
We also found that the analysis of the multi dimensional field 
$q_t(x)$ by the CNN does not improve the accuracy, which suggests
that our CNN fails in capturing useful structures
in the multi dimensional space.

The lattice simulations of this study are in part carried out
on OCTOPUS at the Cybermedia Center, Osaka University.
The NNs are implemented by Chainer framework, and are in part
trained on Google Colaboratory.
This work was supported by JSPS KAKENHI Grant Numbers~17K05442 and
19H05598.

\bibliographystyle{JHEP}
\bibliography{topol_ref}

\end{document}